\documentclass{PoS}

\title{Properties of canonical fermion determinants with a fixed quark number}

\ShortTitle{Determinants with fixed quark number}

\author{Julia Danzer\thanks{This work is supported by FWF DK
W 1203.}\\
        Institute of Physics, University of Graz, Austria\\
        E-mail: \email{julia.danzer@uni-graz.at}}

\author{\speaker{Christof Gattringer}\thanks{C.G.~gratefully acknowledges
    support by the Dr.~Heinrich J\"org-foundation.}\\
        Institute of Physics, University of Graz, Austria\\
        E-mail: \email{christof.gatttringer@uni-graz.at}}
	
\author{Ludovit Liptak \\        
	Institute of Physics, Slovak Academy of Sciences, Dubravska cesta 9,
	845 11 Bratislava, Slovakia\\
        and Institute of Physics, University of Graz, Austria\\
	E-mail: \email{Ludovit.Liptak@savba.sk}}

\abstract{\vspace{10mm}Using a dimensional reduction formula for the lattice 
fermion determinant we
study canonical determinants on quenched SU(3) gauge configurations. The
canonical determinants decribe a fixed quark number and we analyze their
properties below and above the transition temperature. 
We find that above $T_c$ the signatures of center symmetry breaking are 
very strongly manifest in the distribution of the canonical determinants 
in the complex plane, and we discuss possible physical implications of this
finding. We furthermore analyze the relative weight of the 
different quark sectors below and above the transition temperature.}

\FullConference{The XXVII International Symposium on Lattice Field Theory - 
LAT2009\\
		 July 26-31 2009\\
		 Peking University, Beijing, China}

\begin{document}

\section{Introductory comments about canonical determinants}
Canonical fermion determinants $\det[D]^{(q)}$, which describe a fixed number
$q$ of quarks, are conceptionally interesting objects. On the lattice they may
be obtained from the usual grand canonical determinant $\det[D(\mu)]$ through
a Fourier transform with respect to imaginary chemical potential $\mu$:
\begin{equation}
{\det}[D]^{(q)} \; = \; \frac{1}{2\pi} \int_{-\pi}^\pi \!\! d \varphi  \,
e^{-i q \varphi}\, \det[D(\mu = i\varphi/\beta)] \; .
\label{DQfourierdef} 
\end{equation} 
Here $\beta$ is the inverse temperature which is given by the (periodic)
temporal extent of the lattice.  $\varphi$ is the angle that parameterizes the
imaginary chemical potential $\mu = i\varphi/\beta$. The individual canonical
determinants  $\det[D]^{(q)}$ appear as coefficients in the fugacity expansion
of the grand canonical determinant $\det[D(\mu)]$,
\begin{equation}
\det[D(\mu)] \; = \; \sum_q \, 
e^{\mu q \beta} \, \det[D]^{(q)} \; .
\label{fugexpand}
\end{equation}
Thus the representation with the canonical determinants is equivalent to the
grand canonical formulation. In recent years several numerical simulations in
the canonical formalism may be found in the literature   \cite{canonical1} --
\cite{canonical3}, and were reviewed at this conference \cite{canonical4}.

Canonical determinants do not only provide an alternative approach to lattice
simulations with finite density, but also have interesting physical
properties \cite{faber}. 
The grand canonical determinant $\det[D(\mu)]$ is a gauge
invariant object and thus is a sum of products of closed loops which are
dressed with link variables. The chemical potential is equivalent to a
temporal fermionic boundary condition $\exp(i\varphi)$. This phase at the
boundary is seen by the loops that wind around the compact time direction
according to their total winding number and gives rise to a phase factor
$\exp(i\varphi k)$ for a loop that winds $k$-times. The Fourier integral
(\ref{DQfourierdef}) over this boundary condition projects the grand canonical
determinant to only those loops which have a net winding number of $k=q$ (see,
e.g., \cite{GaLi} for a more detailed discussion of these relations). The
fugacity expansion (\ref{fugexpand}) thus may also be viewed as an expansion
in terms of winding numbers of loops. 

In a similar way all gauge invariant objects may be decomposed into sectors of
loops with fixed winding number. This has been discussed for the chiral
condensate, where it is hoped that the "dual chiral condensate", defined as
the sector of the chiral condensate with winding 1, might help to understand a
possible relation between chiral symmetry restoration and deconfinement
\cite{dualcond} -- \cite{dualcond3}.

\section{Center symmetry} 
The canonical determinants have simple transformation properties under center
transformations, where all temporal gauge links $U_4(\vec{x},t_0)$ at a fixed
time argument $t_0$ are multiplied with an element $z$ of the center of the
gauge group, i.e., $U_4(\vec{x},t_0) \rightarrow z \, U_4(\vec{x},t_0)$. For
the case of gauge group SU(3) the $z$ are the phases $z = 1, \exp(\pm i
2\pi/3)$. The gauge action and the gauge measure are invariant under center
transformations. Consequently the expectation functional $\langle
.. \rangle_G$ for the evaluation of observables in pure gauge theory is
invariant under center transformations, as long as the center symmetry is not
broken spontaneously. However, such a spontaneous breaking of the center
symmetry takes place at high temperatures \cite{centerbreaking}. 

Observables may be classified with respect to their symmetry properties under
center transformations. A simple  example is the Polyakov loop $P$ which is
the trace over a temporal gauge transporter that winds in a straight line once
around compact time. As it winds once, it sees exactly one of the link
variables $U_4(\vec{x},t_0)$ which are transformed with the center element $z$
and we conclude that $P$ transforms as $P \rightarrow zP$. Since it transforms
non-trivially, the Polyakov may be used as an order parameter for the breaking
of center symmetry. Below the critical temperature $T_c$ its expectation value
vanishes, while above $T_c$ this expectation value is finite. 

\begin{figure}[t!]
\begin{center}
\includegraphics[height=45mm,clip]{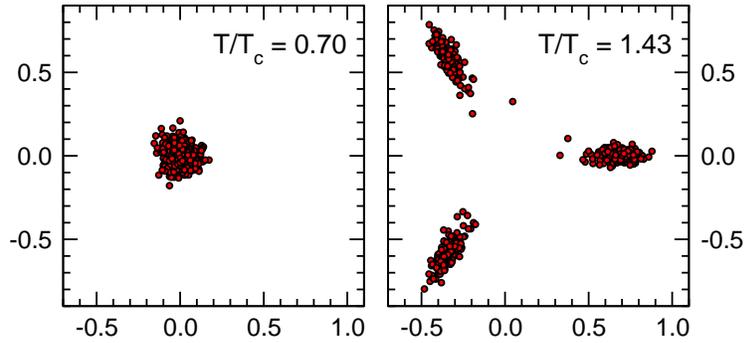}
\end{center}
\caption{Scatter plots of Polyakov loop values in the complex 
plane for low (lhs.) and high temperature.}
\label{Ploopscatter}
\end{figure}

The behavior of the Polyakov loop is illustrated in Fig.~\ref{Ploopscatter}
where we show scatter plots of the Polyakov loop in the complex plane. The
data are from 500 quenched gauge configurations on $8^3 \times 4$ lattices
generated with the L\"uscher-Weisz gauge action \cite{luweact}. We show two
ensembles with temperatures of $T/T_c = 0.7$ (lhs.\ plot) and $T/T_c = 1.43$
(rhs.) according to the scale setting \cite{scale} with the Sommer
parameter. Obviously for low temperatures the values of $P$ are compatible
with zero, while they are non-vanishing above $T_c$, where the center symmetry
is broken. We note that a true spontaneous breaking can happen only at
infinite spatial volume. In that case the system will spontaneously select one
of the three "islands" in the complex plane and only populate this one island.

Similar to the Polyakov loop we also may obtain the transformation properties
of our canonical determinants  $\det[D]^{(q)}$. We have already observed that
they consist of loops with a net winding number of $q$ around compact
time. Thus they have a net number of $q$ crossings of the time slice $t_0$
where the center transformation acts and consequently transform as
\begin{equation}
\det[D]^{(q)} \; \; \longrightarrow \; \; z^{\,q} \; 
\det[D]^{(q)} \; \; = \; \; z^{\,q \, \mbox{mod} \, 3} 
\; \det[D]^{(q)} \; ,
\label{DQcentertrafo}
\end{equation}
where in the last step we have used that $z = 1, \exp(\pm i 2\pi/3)$. 

The canonical partition sums $Z^{(q)}$ for a fixed quark number $q$ are
obtained as the expectation values of the canonical determinants using the
pure gauge theory expectation functional $\langle .. \rangle_G$. As long as
the center symmetry is unbroken, we find
\begin{equation}
Z^{(q)} \; = \; \langle \det[D]^{(q)} \rangle_G \; 
\stackrel{\mbox{c.u.~!}}{ = } \; z^q \,  
\langle \det[D]^{(q)} \rangle_G \; \; \; \Longrightarrow \; \; \; 
Z^{(q)} \, = \, 0 \;\; \mbox{for} \; \;
q \, \mbox{mod} \, 3 \, \neq \, 0 \; .
\label{triality}
\end{equation}
Thus in the low temperature phase, where the center symmetry is unbroken, the
canonical partition sums $Z^{(q)}$ are non-vanishing only for quark sectors
with vanishing triality, i.e., when $q$ is a multiple of 3. The fact that the
center symmetry is unbroken is crucial for the argument in (\ref{triality}),
and we marked the step where we use that property by "c.u.~!" for "center
unbroken". We remark that the transformation properties (\ref{DQcentertrafo})
may be combined with the center properties of observables to derive selection
rules for observables in the center symmetric phase \cite{Ploopcanonical},
\cite{GaLi}. One finds that the total triality of an observable multiplied
with a canonical determinant has to vanish for a non-vanishing contribution
in the center symmetric phase.

The situation is different in the deconfined high temperature phase where
center symmetry is broken spontaneously.  The argument used in
(\ref{triality}) may no longer be applied, and $Z^{(q)}$ can be non-vanishing
also for $q$ mod $3 \neq 0$. In other words, above $T_c$ also canonical
determinants $\det[D]^{(q)}$ with non-vanishing triality can have a
non-vanishing expectation value. 

\begin{figure}[t]
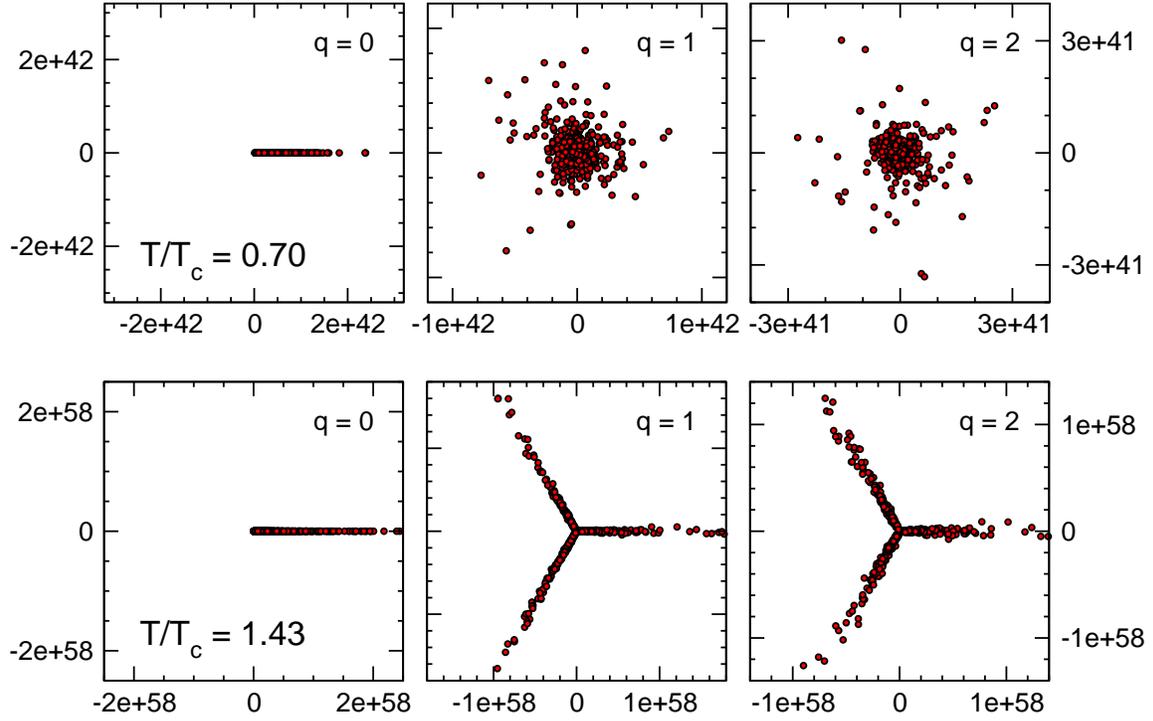

\begin{center}
\includegraphics[height=45mm,clip]{DQ_scatter_8x4_b740.eps}
\vskip5mm
\includegraphics[height=45mm,clip]{DQ_scatter_8x4_b820.eps}
\end{center}
\caption{Scatter plots of the canonical determinants 
$\det[D]^{(q)}, q = 0,1,2$ (left to right) in the complex 
plane. The top row is for $T/T_c = 0.70$, the bottom for $T/T_c = 1.43$.}
\label{DQscatter}
\end{figure}

In Fig.~\ref{DQscatter} we study the behavior of the canonical determinants
$\det[D]^{(q)}$ below (top row of plots) and above $T_c$ (bottom) for quark
numbers $q = 0,1,2$ (plots from left to right). We show scatter plots of the
values of the canonical determinants in the complex plane. The canonical
determinants were evaluated as described in \cite{GaLi}, using a dimensional
reduction formula for determinants \cite{DaGa}. The lattice volume is  $8^3
\times 4$, the bare quark mass parameter in the fermion determinant is set to
$m$ = 100 MeV, and the statistics is 500 configurations for $T < T_c$ (top
row) and 1000 configurations for $T > T_c$ (bottom).

For the zero triality case $q=0$ (left column of plots) the values of
$\det[D]^{(0)}$ fall on the positive half of the real axis and thus give rise
to a positive expectation value $Z^{(0)}$ both below and above $T_c$. For the
non-vanishing triality sectors with $q = 1, 2$ (center and rhs.~plots) the
properties below and above $T_c$ are drastically different. Below $T_c$ the
values of $\det[D]^{(q)}, q = 1,2$, scatter around the origin in a spherically
symmetrical distribution. Above $T_c$ the distribution is rather different and
we observe the center symmetry pattern familiar from the Polyakov loop as
shown in Fig.~\ref{Ploopscatter}. We stress that for the canonical
determinants the pattern is even much cleaner than for the Polyakov loop
(Both, Fig.~\ref{Ploopscatter} and Fig.~\ref{DQscatter} were made with the
same $8^3 \times 4$ ensembles.). 
 
Let us at this point address again the role of the infinite volume, which is
necessary for a spontaneous breaking of the center symmetry. The high
temperature data in Figs.~\ref{Ploopscatter} and \ref{DQscatter} show the
star-like pattern  characteristic for the center broken phase. However, all
three center sectors are populated equally and a naive averaging over all
points in the scatter plots would give a vanishing expectation value for the
Polyakov loop $P$, as well as the canonical determinants $\det[D]^{(q)}, q =
1,2$ at all temperatures -- a truly uninteresting outcome. To obtain the
physically relevant result one must take into account that in the limit of
infinite spatial volume, the system selects spontaneously only one of the
three sectors, with the other two remaining empty. In a simulation on a finite
lattice this may be taken into account by considering above $T_c$ the absolute
values of symmetry breaking observables, i.e., $\langle |P| \rangle_G$ and
$\langle |\det[D]^{(q)}| \rangle_G$. 

\begin{figure}[t]
\begin{center}
\includegraphics[height=147mm,clip]{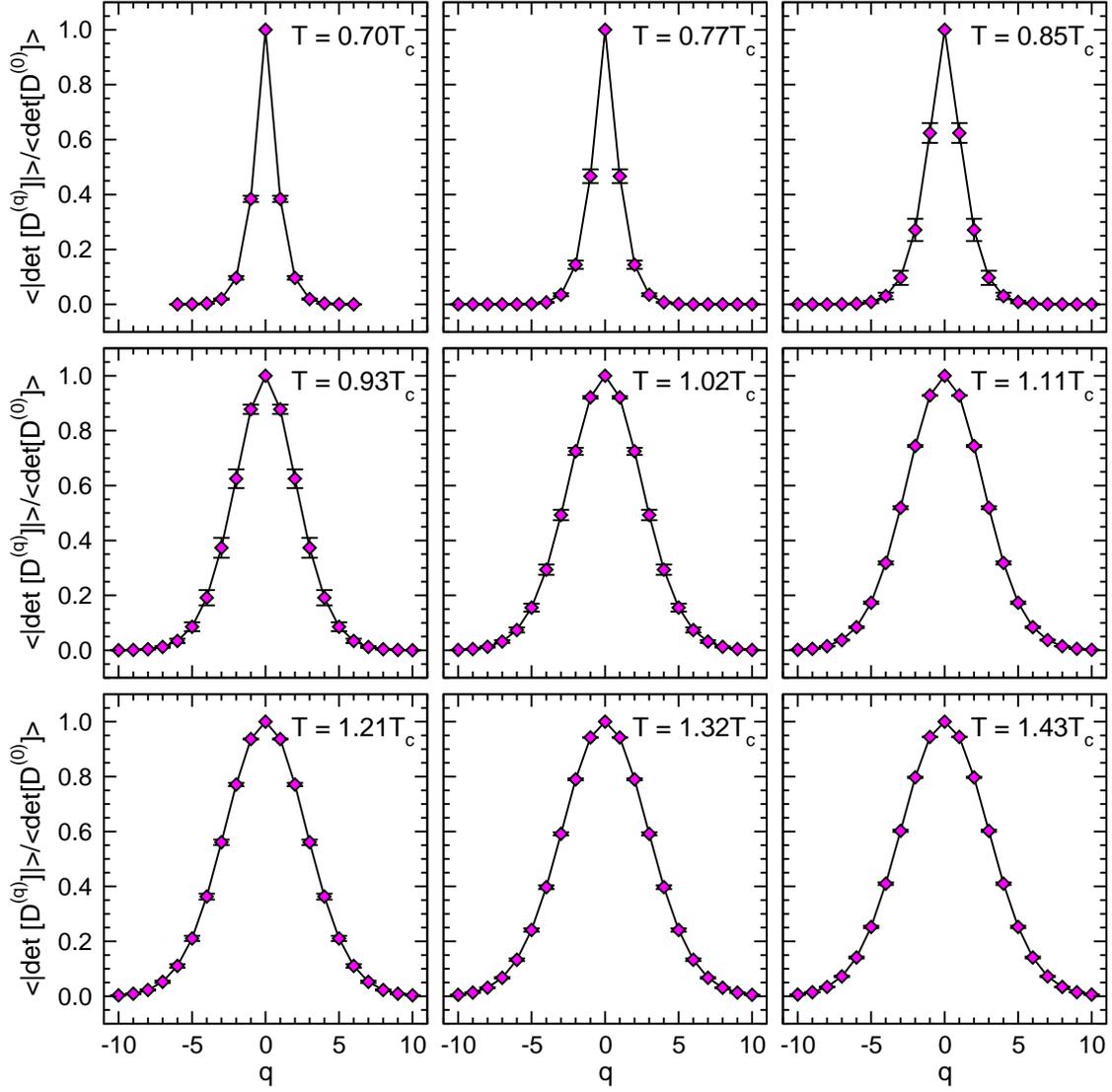}
\end{center}
\caption{The distribution of $\langle | \det[D]^{(q)} | \rangle_G / \langle
  \det[D]^{(0)} \rangle_G$ as a function of the quark number $q$ for different
  values of the temperature.}
\label{DQdist}
\end{figure}

\section{Distribution of the quark sectors}
Let us now  study in more detail how the canonical  determinants behave in the
high temperature phase. In particular how sectors with different quark numbers
behave relative to each other. As already discussed in the last section, above
$T_c$, where  the center symmetry is broken,  also the sectors with  $q$ mod 3
$\neq 0$ can  have non-vanishing expectation values. As  also addressed there,
one has to average  the absolute value of the determinants, as  long as one is
on a  finite volume where the center  symmetry cannot be broken  and all three
center sectors are populated equally (if a proper Monte Carlo update is used).

In Fig.~\ref{DQdist} we show the distribution $\langle | \det[D]^{(q)} |
\rangle_G / \langle \det[D]^{(0)} \rangle_G$ as a function of $q$, i.e., we
normalize with respect to the trivial sector with $q = 0$. The data are for
lattice size $8^3 \times 4$, a quark mass of $m = 100$ MeV, and a statistics
of 100 configurations. We show the results for different temperatures,
ranging from $T  = 0.70 T_c$ to  $T  = 1.43 T_c$

Below $T_c$ the distribution of the
absolute value of the canonical determinants shows a Gaussian type of
behavior with a rather narrow width. This width is increasing with temperature.
Above $T_c$ the distribution remains Gaussian, but the width does not seem to
grow any longer with temperature. We remark, however, at this point, that we
work with a lattice size fixed in lattice units ($8^3 \times 4$), and change
the temperature by varying the gauge coupling. Thus increasing the temperature 
also shrinks the spatial volume. This effect could mask a further widening of
the distribution above $T_c$, but even if such widening persists, it is much
smaller than the effect seen below $T_c$. A detailed finite volume analysis of
the quark distribution must be left for future studies. 

We conclude with discussing an important consistency check: Once the canonical
determinants are known, one can try to sum up the fugacity expansion and
compare this sum to the grand canonical fermion determinant. The plots in 
Fig.~\ref{DQdist} show that the canonical determinants quickly approach zero
as $|q|$ increases, and a truncation of the fugacity expansion seems
justified. We implemented such a test and summed the fugacity expansion,
taking into account terms with $|q|$ up to typically values of
30-50. For moderate chemical potential of up to $a\mu \sim 0.1$ we found
excellent agreement (relative error less than $10^{-4}$) between the fugacity
sum and the grand canonical result, showing that the determination of the
lowest canonical determinants is sufficiently accurate. For larger values of
the chemical potential higher terms start to contribute which would have to 
be evaluated with higher accuracy.

\end{document}